\documentclass{ws-ijmpc_rev}
\usepackage{graphicx}

\begin{document}

\markboth{Hiroshi Koibuchi and Andrey Shobukhov}
{Monte Carlo simulations of Landau-Ginzburg model for membranes}

\catchline{}{}{}{}{}

\title{Monte Carlo simulations of Landau-Ginzburg model for membranes
}

\author{Hiroshi Koibuchi
}

\address{Department of Mechanical and Systems Engineering, Ibaraki National College of Technology, Nakane 866 Hitachinaka, Ibaraki 312-8508, Japan
\\koibuchi@mech.ibaraki-ct.ac.jp
}
 
\author{Andrey Shobukhov}

\address{Faculty of Computational Mathematics and Cybernetics, Lomonosov Moscow State University, 
119991, Moscow, Leninskiye Gory, MSU, 2-nd Educational Building, Russia
}

\maketitle

\begin{history}
\received{January 2014}
\revised{January 2014}
\end{history}

\begin{abstract}
The Landau-Ginzburg (LG) model for membranes is numerically studied on triangulated spheres in ${\bf R}^3$.  The LG model is in sharp contrast to the model of Helfrich-Polyakov (HP). The reason for this difference is that the curvature energy of the LG (HP) Hamiltonian is defined by means of the tangential (normal) vector of the surface. For this reason the curvature energy of the LG model includes the in-plane bending or shear energy component, which is not included in the curvature energy of the HP model. From the simulation data, we find that the LG model undergoes a first-order collapse transition. The results of the LG model in the higher dimensional spaces ${\bf R}^d (d>3)$  and on the self-avoiding surfaces in  ${\bf R}^3$ are presented and discussed. We also study the David-Guitter (DG) model, which is a variant of the LG model, and find that the DG model undergoes a first-order transition. It is also found that the transition can be observed only on the homogeneous surfaces, which are composed of almost uniform triangles according to the condition that the induced metric ${\partial_a {\bf r}}\cdot{\partial_b {\bf r}}$ is close to $\delta_{ab}$. 

\end{abstract}

\ccode{PACS Nos.:  64.60.-i, 68.60.-p, 87.10.-e, 87.16.D-}

\section{Introduction}\label{introduction}

All numerical studies that have been performed so far for understanding the shape transformations of membranes ignore the Landau-Ginzburg (LG) model \cite{PKN-PRL1988,Bowick-SMMS2004}, which is based on the usage of the surface tangential vector, and apply the Helfrich-Polyakov (HP) model \cite{HELFRICH-1973,POLYAKOV-NPB1986}, which is defined by means of the normal vector to the surface in ${\bf R}^3$ \cite{KANTOR-NELSON-PRA1987,Kantor-SMMS2004,Gompper-Kroll-SMMS2004}. The HP model includes the spring and beads model, of which the curvature energy is defined by the normal vector \cite{KANTOR-NELSON-PRA1987}. To the contrary, the order parameter of the LG model is the tangential vector $\partial_a {\bf r}(a\!=\!1,2)$ of the surface, the position of which is denoted by ${\bf r}$. Therefore, the definition of the HP model is completely different from that of the LG model, and for this reason it is still unclear whether or not the numerical result of the HP model should be consistent with the theoretical prediction of the LG model. In fact, numerical simulations of the LG model are yet to be performed. 

We should mind that the curvature energy in the LG Hamiltonian consists of the out-of-plane and in-plane bending components. In addition,  an in-plane shear energy component is also included in the energy term of the fourth power of $\partial_a {\bf r}$ in the LG Hamiltonian. In contrast, the HP Hamiltonian has the extrinsic curvature energy without the in-plane shear energy component. For this reason, the LG model is in sharp contrast to the HP model. Moreover, we know that the existence of in-plane shear energy influences the order of the crumpling transition in the case of meshwork models \cite{Endo-Koibuchi-EPJB2008}. Therefore, we consider it is a nontrivial problem to find out whether or not the order of the transition of the LG model is identical with that of the HP model. 

It should also be minded that the order parameter $\partial_a {\bf r}$ forms the induced metric $\partial_a {\bf r}\cdot\partial_b {\bf r}$ of the surface. This metric is closely related to the shape of triangles in the discrete model on triangulated surfaces, because $\partial_a {\bf r}$ just corresponds to the edge length of triangles. The regular triangle corresponds to the Euclidean metric $\delta_{ab}$ which is a special case of $\partial_a {\bf r}\cdot\partial_b {\bf r}$. 
 Moreover, the anisotropic shape transformation is a direct consequence of the anisotropic shape of triangles in the triangulated surface model \cite{Koibuchi-EPJB2007}. Thus, the model of David-Guitter (DG) is very interesting because it demonstrates the dependence of the transition on the shape of triangles \cite{David-Guitter-1988EPL,FDavid-SMMS2004}. Indeed, the DG Hamiltonian includes the energy term of the strain tensor $u_{ab}\!=\!\partial_a {\bf r}\cdot\partial_b {\bf r}\!-\!\delta_{ab}$, which measures a difference between the induced metric and the Euclidean metric.  

In this paper, we numerically study the LG and DG models on triangulated surfaces. Our aim is to find out whether the first-order transition is seen in the models or not and to find the dependence of the transition on the shape of triangles. Both the  self-avoiding and phantom surface models are studied. Here the phantom model is the model of the surface that is allowed to self-intersect. 

This paper is organized as follows. In Section \ref{LG-model}, we study the LG model. The continuous LG model is introduced and the mean field description of the model is reviewed in Subsection \ref{LG-continuous}. The discretization of the LG model on triangulated surfaces is described in detail in Subsection \ref{LG-discrete}, and the MC results are shown in Subsection \ref{LG-Monte}, where the first-order transition is confirmed. The MC results of the LG model in higher dimensions and those of the self-avoiding surfaces are briefly presented and discussed in Subsection \ref{LG-SA-Highdim}. In Section \ref{DG-model}, we study the DG model. The continuous DG model is introduced, and a relation between the strain tensor and the surface metric is mentioned in Subsection \ref{DG-continuous}. The discrete DG model is described in Subsection \ref{LG-discrete}, and the MC results are presented in Subsection \ref{DG-Monte}. We summarize the results in Section \ref{Conclusion}.

\section{Landau-Ginzburg model for membranes}\label{LG-model}
\subsection{Continuous Landau-Ginzburg Hamiltonian}\label{LG-continuous}
Let ${\bf r}\!=\!(X,Y,Z)$ be the three-dimensional position of a surface embedded in ${\bf R}^3$. The Landau-Ginzburg Hamiltonian for membranes is given by \cite{PKN-PRL1988,Bowick-SMMS2004}
\begin{eqnarray}
\label{LG-Hamiltonian}
S_{\rm LG}({\bf r})&=&\frac{t}{2}\int d^2x \left(\partial_a {\bf r} \right)^2 
+ \frac{\kappa}{2} \int d^2x\left(\partial^2 {\bf r} \right)^2  \nonumber \\
&+& u \int d^2x\left(\partial_a {\bf r}\cdot\partial_b {\bf r} \right)^2
+ v \int d^2x\left(\partial_a {\bf r}\cdot\partial_a {\bf r} \right)^2  \nonumber \\
&=&tS_1 +\kappa S_2 + uS_3+vS_4. 
\end{eqnarray}
The variable $x\!=\!(x_1,x_2$) denotes a local coordinate of the surface. The Hamiltonian $S_{\rm LG}({\bf r})$ should be written as $S_{\rm LG}(\partial {\bf r})$ since $\partial {\bf r}$ is the order parameter,  however, we here write $S_{\rm LG}$  as $S_{\rm LG}({\bf r})$ because ${\bf r}$ is the dynamical variable of the model. The first term $S_1$ is identical to the Gaussian bond potential in the HP model and represents the in-plane tensile elasticity, and the coefficient $t$ is the microscopic surface tension. The second term $S_2$ is the curvature energy with the bending rigidity $\kappa$. 

Here we comment on the difference between this $S_2$ and that of the HP model. This $S_2$ can also be written as $S_2\!=\!\int \sqrt{g}d^2x \left(g^{ab} {\partial_a\partial_b {\bf r}}\right)^2\!=\!\int \sqrt{g}d^2x \left(g^{ab} {\partial_a {\bf e}_b}\right)^2$, where $g$ is the determinant of the metric tensor $g_{ab}$ of the surface, $g^{ab}$ is its inverse, and the tangential vector ${\bf e}_a$ is defined by ${\bf e}_a\!=\!\partial_a {\bf r}\!=\!\partial {\bf r}/\partial x_a$. Note that ${\bf e}_a$ is written as ${\bf t}_a$ in Ref. \refcite{FDavid-SMMS2004}. Since ${\partial_a {\bf e}_b}\!=\!\Gamma_{ab}^k{\bf e}_k\!+\!K_{ab}{\bf n}$ (Gauss's equation), $S_2$ can be rewritten as $S_2\!=\!\int \sqrt{g}d^2x g^{ij}g^{kl} \left(\Gamma_{ij}^a\Gamma_{kl}^bg_{ab}\!+\!K_{ij}K_{kl}\right)$, where $\Gamma_{ij}^a(=\!g^{ab}\Gamma_{ibj})$ is the Christoffel symbols of the second kind, which is called the affine connection \cite{FDavid-SMMS2004}, $\Gamma_{ibj}\!=\!{\bf e}_b\cdot{\partial_j {\bf e}_i}$ is the Christoffel symbols of the first kind,  and $K_{ij}\!=\!{\partial_j {\bf e}_i}\cdot{\bf n}$ is the second fundamental form of the surface. The first term $\int \sqrt{g}d^2x g^{ij}g^{kl} \Gamma_{ij}^a\Gamma_{kl}^bg_{ab}$ and the second term $\int \sqrt{g}d^2x g^{ij}g^{kl} K_{ij}K_{kl}$ in this $S_2$ are considered to be the in-plane and out-of-plane bending energies, respectively.  
Thus, recalling that the second term $\int \sqrt{g}d^2x g^{ij}g^{kl} K_{ik}K_{jl}\!=\!\int \sqrt{g}d^2x K^j_{k}K^k_{j}$ is just the same as the bending energy of the HP model, we find that $S_2$ in Eq. (\ref{LG-Hamiltonian}) of the LG model is different from that of the HP model \cite{POLYAKOV-NPB1986}. Note also that $\int \sqrt{g}d^2x K^j_{k}K^k_{j}$ can also be written as $\int \sqrt{g}d^2x K^j_{k}K^k_{j}\!=\!\int\sqrt{g}d^2x g^{ij}{\partial_i {\bf n}}\cdot{\partial_j {\bf n}}$ since $g_{ab}$ is given by $g_{ab}\!=\!\partial_a {\bf r}\cdot\partial_b {\bf r}$. This final expression represents the out-of-plane bending energy. 

The third term $S_3$ in $S_{\rm LG}$ of Eq. (\ref{LG-Hamiltonian}) also has an in-plane shear energy component, while $S_4$ has not. Thus, recalling that an in-plane shear energy influences the  transition in the meshwork models \cite{Endo-Koibuchi-EPJB2008}, we find it is non-trivial that the phase structure of the LG model is identical with that of the HP model, in which a first-order crumpling transition is observed \cite{KD-PRE2002,Koibuchi-PRE2004,Koibuchi-PRE2005}.

Assuming $g_{ij}\!=\!\delta_{ij}$, we can explicitly write down the Hamiltonians of Eq. (\ref{LG-Hamiltonian}):
\begin{eqnarray}
\label{cont_LG}
S_{\rm LG}  &=& tS_1 +\kappa S_2 + uS_3+vS_4,  \nonumber \\
S_1 &=& \frac{1}{2}\int d^2x \left[\left(\partial_1 {\bf r} \right)^2+\left(\partial_2 {\bf r} \right)^2\right], \nonumber \\
S_2 &=& \frac{1}{2} \int d^2x\left(\partial_1^2 {\bf r}+\partial_2^2 {\bf r} \right)^2  \nonumber \\
&=&\frac{1}{2} \int d^2x\left(\partial_1^2 {\bf r}\cdot\partial_1^2 {\bf r}+\partial_2^2 {\bf r}\cdot\partial_2^2 {\bf r}+2\partial_1^2 {\bf r}\cdot\partial_2^2 {\bf r}\right),   \\
S_3 &=&   \int d^2x\left[\left(\partial_1 {\bf r}\cdot\partial_1 {\bf r} \right)^2+\left(\partial_2 {\bf r}\cdot\partial_2 {\bf r} \right)^2+2\left(\partial_1 {\bf r}\cdot\partial_2 {\bf r} \right)^2\right],\nonumber \\
S_4 &=& \int d^2x\left[\left(\partial_1 {\bf r}\cdot\partial_1 {\bf r} \right)^2+\left(\partial_2 {\bf r}\cdot\partial_2 {\bf r} \right)^2 + 2\left(\partial_1 {\bf r}\right)^2\left(\partial_2 {\bf r}\right)^2 \right]. \nonumber
\end{eqnarray}
The expression ${\partial ^2}$ in $S_2$ is the Laplacian, and ${\partial ^2 {\bf r}}$ is formally written as $g^{ab}{\partial_a\partial_b {\bf r}}$ with the inverse metric $g^{ab}$ on the surface. In the case of $g_{ab}\!=\!\delta_{ab}$,  ${\partial ^2 {\bf r}}$ is written as $g^{ab}{\partial_a\partial_b {\bf r}}\to \partial_1^2 {\bf r}\!+\!\partial_2^2 {\bf r}$, and we obtain the expression in Eq. (\ref{cont_LG}). Note also that the third term in $S_3$ plays a role of in-plane shear energy, while all terms in $S_4$ are in-plane tensile energy, as mentioned above. 

Finally in this subsection, we present the mean field analysis of the model introduced in Ref. \refcite{Bowick-SMMS2004}. Using the Monge representation, the variable ${\bf r}$ can be expanded as 
\begin{eqnarray}
{\bf r}=\left(\zeta {\bf x}+{\bf u}({\bf x}), h({\bf x})\right),\quad {\bf x}=(x_1,x_2)
\end{eqnarray}
where  ${\bf u}({\bf x})$ and $h({\bf x})$ are small fluctuations of ${\bf r}$ around $(\zeta {\bf x},0)$, which gives the mean filed $\partial {\bf r}\!=\!(\zeta ,\zeta ,0)\in {\bf R}^3$. From this expansion, the potential $V$ is derived in  Ref. \refcite{Bowick-SMMS2004} as the following function of $\zeta$: 
\begin{eqnarray}
\label{mean_field}
V(\zeta)=2\zeta ^2\left(\frac{t}{2}+(u+2v)\zeta^2\right).
\end{eqnarray}
Thus, we understand that $V(\zeta)$ has a double minima as a function of $\zeta$ if $t<0$, and that the variable $\zeta$ may be regarded as the  magnitude of the mean field up to a numerical factor. Therefore from the Landau theory of phase transitions it follows that the model undergoes a continuous crumpling transition if $t<0$. However, this prediction is correct only if the fluctuations around the mean field are very small compared to the mean field itself. Moreover, the crumpling transition is characterized by large fluctuations of the surface. Therefore, it is quite natural that the continuous nature of the transition predicted by the mean field analysis is not always correct. For this reason, we consider it is worthwhile to find out numerically whether the LG model undergoes a first-order transition or not.

\subsection{Discrete Landau-Ginzburg Hamiltonian}\label{LG-discrete}
\begin{figure}[h]
\centering
\includegraphics[width=11.5cm]{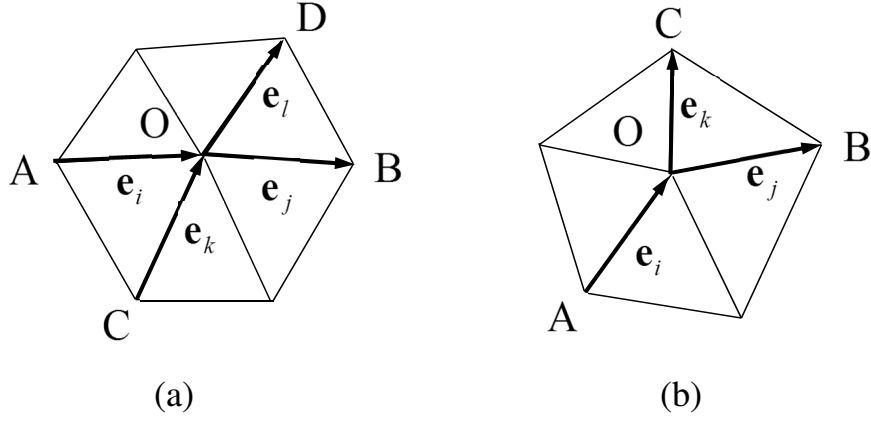}  
\caption{Tangential (or edge) vectors along the edges of triangles in (a) a hexagon and (b) a pentagon. 
 The vectors  ${\bf e}_i$ and ${\bf e}_j$ are on the diagonal line (AOB) of the hexagonal lattice, and those ${\bf e}_k$ and ${\bf e}_l$ are also on another diagonal line (COD). The sum $\sum_{ij}$ in the first term of $S_2$ in Eq. (\ref{disc_LG}) denotes the sum over all three possible diagonal lines $ij$ on a hexagon, while $\sum_{(ij),(kl)}$  in the second term denotes the sum over all three possible coordinates  ${(ij),(kl)}$ on the hexagon.
 } 
\label{fig-1}
\end{figure}
 The variables $(x_1,x_2)$ are the local coordinates on the surface, and the Monge gauge is not always assumed henceforth. The discrete LG Hamiltonian on a triangulated surface is given by
\begin{eqnarray}
\label{disc_LG} 
&&S_{\rm LG}=tS_1 +\kappa S_2 + uS_3+vS_4,  \nonumber \\
&&S_1=\frac{2}{3}\sum_{ij} \left({\bf r}_i-{\bf r}_j\right)^2=\frac{2}{3}\sum_{j} {\bf e}_{j}^2, \nonumber \\
&&S_2=\frac{1}{3}\sum_{ij}\left({\bf e}_i- {\bf e}_j\right)^2+\frac{1}{3}\sum_{(ij),(kl)}\left({\bf e}_i- {\bf e}_j\right)\cdot\left({\bf e}_k- {\bf e}_l\right),  \\
&&S_3=\frac{2}{3}\sum_{i=1}^{N_T} \left[\left({\bf e}_1^2\right)^2 + \left({\bf e}_2^2\right)^2 + \left({\bf e}_3^2\right)^2
+\left({\bf e}_1\cdot{\bf e}_2\right)^2 +\left({\bf e}_2\cdot{\bf e}_3\right)^2 +\left({\bf e}_3\cdot{\bf e}_1\right)^2 \right], 
\nonumber \\
&&S_4=\frac{2}{3}\sum_{i=1}^{N_T} \left[\left({\bf e}_1^2\right)^2 + \left({\bf e}_2^2\right)^2 + \left({\bf e}_3^2\right)^2
+\left({\bf e}_1^2\right)\left({\bf e}_2^2\right) +\left({\bf e}_2^2\right)\left({\bf e}_3^2\right) +\left({\bf e}_3^2\right)\left({\bf e}_1^2\right) \right].\nonumber
\end{eqnarray} 
Since $S_2$ in Eq. (\ref{disc_LG}) is defined by a derivation of edge vectors, the discretization of this derivation can not be performed on a single triangle; this is in sharp contrast to the discretization of the other energies composed of only edge vectors. In  Fig. \ref{fig-1}(a), the diagonal lines AOB and COD play the role of local coordinates $(x_1,x_2)$ in the hexagon such that $x_1$ is constant on COD and $x_2$ is constant on AOB. Using these local coordinates, the edge vector ${\bf e}_1\!:=\!\partial {\bf r}/\partial x_1$ at O can be replaced by
\begin{equation}
\label{bond-vect}
{\bf e}_1={\bf r}({\rm B})-{\bf r}({\rm O}), 
\end{equation}
which is written as ${\bf e}_j$ in Fig. \ref{fig-1}(a). Thus, $\partial_1^2 {\bf r}$ and $\partial_2^2 {\bf r}$ can be replaced by ${\bf e}_j\!-\!{\bf e}_i$ and ${\bf e}_l\!-\!{\bf e}_k$, respectively. Therefore, the square of Laplacian $\left( \partial ^2 {\bf r} \right)^2\!=\!(\partial_1^2 {\bf r}+\partial_2^2 {\bf r} )^2$ is replaced by $\left({\bf e}_j\!-\!{\bf e}_i\right)^2\!+\!\left({\bf e}_l\!-\!{\bf e}_k\right)^2\!+\!2\left({\bf e}_j\!-\!{\bf e}_i\right)\cdot\left({\bf e}_l\!-\!{\bf e}_k\right)$ in the local coordinates of the hexagon in Fig. \ref{fig-1}(a). Note that we have three independent coordinates $(x_1,x_2)$ in the hexagon, because three different diagonal lines are possible in it (Fig. \ref{fig-1}(a)).  For this reason,  $\left( \partial ^2 {\bf r} \right)^2$ has three different sets of discretization corresponding to those three coordinates. Thus, summing over all different sets of discretization, we define a discretization of $\partial^2 {\bf r}$ in the hexagon. This summation is performed in all hexagons, and hence we have $S_2$ in Eq. (\ref{disc_LG}) as a discrete bending energy corresponding to the continuous one $(1/2)\int dx^2\left( \partial ^2 {\bf r} \right)^2$. The reason why the factor $1/3$ is included in the expression is because every vertex is assumed to be the center of hexagon, and therefore the summation is triply duplicated. In $S_2$, $\sum_{ij}$ denotes the sum over the three different diagonal lines, and $\sum_{(ij),(kl)}$ denotes the sum over the corresponding local coordinates on the hexagon.  

On a pentagon, $\partial_i^2 {\bf r}$ can be replaced by ${\bf e}_j\!-\!{\bf e}_i$ and ${\bf e}_k\!-\!{\bf e}_i$ (Fig. \ref{fig-1}(b)). The vectors ${\bf e}_j$ and ${\bf e}_k$  form a diagonal line together with ${\bf e}_i$ on the pentagon. Since we have those five different diagonal lines on a pentagon, the quantity such as $({\bf e}_j\!-\!{\bf e}_i)^2$ contributes to the summation in $S_2$ with the weight of $1/2$ on a pentagon.

It is easy to see that $S_3$ and $S_4$ in Eq. (\ref{disc_LG}) are the discretizations of the continuous $S_3$ and $S_4$ in Eq. (\ref{cont_LG}). The symbol $N_T$ in $S_3$ and $S_4$ denotes the total number of triangles. Note that $S_3$ and $S_4$ in Eq. (\ref{disc_LG}) include only the first order derivatives of ${\bf r} $, and for this reason $S_3$ and $S_4$ are discretized on a single triangle. This is in a sharp contrast with the case of discretization for $S_2$ in Eq. (\ref{disc_LG}) as mentioned above.

The partition function is given by
\begin{equation}
\label{part-func-LG}
Z=\int^\prime \prod_i d{\bf r}_i \exp \left( -S_{\rm LG}\right),
\end{equation}
where the prime in $\int^\prime \prod_i d{\bf r}_i$ denotes that the center of the mass of surface is fixed at the origin of ${\bf R}^3$. From the scale invariance of $Z$ \cite{WHEATER-JP1994}, we have
\begin{equation}
\label{S1234}
S_1^\prime/N=3/2, \quad {\rm where}  \quad S_1^\prime=t S_1 + \kappa S_2 + 2u S_3 +2v S_4.
\end{equation}
Indeed, the replacement of the variable ${\bf r}\!\to\!{\bf r}^\prime\!=\!\alpha {\bf r}$ changes $Z$ so that $Z(\alpha)\!=\!\alpha^{3N-1}\int^\prime \prod_i d{\bf r}_i \exp \left[ -\left(t \alpha^2S_1 \!+\! \kappa \alpha^2S_2 \!+\! u \alpha^4S_3 \!+\!v \alpha^4S_4\right)\right]$. Since the integral in $Z$ is invariant under such a variable transformation, we have $\partial Z(\alpha)/ \partial \alpha |_{\alpha=1}$. From this, we obtain $S_1^\prime/N\!=\!3/2$ in Eq. (\ref{S1234}) in the limit of $N\!\to\!\infty$, where $N$ is the total number of vertices.

\subsection{Monte Carlo results}\label{LG-Monte}
The canonical Metropolis  Monte Carlo (MC) technique is used. 
The total number of MC sweeps (MCS) is approximately $2\!\times\! 10^9\!\sim\! 4\!\times\! 10^9$ at the transition region and relatively small ($0.5\!\times\! 10^9\!\sim\! 2\!\times\! 10^9$ ) at the non-transition region. The measurements of data are done every 1000 MCS during the simulations.

\begin{figure}[!t]
\centering
\includegraphics[width=12.5cm]{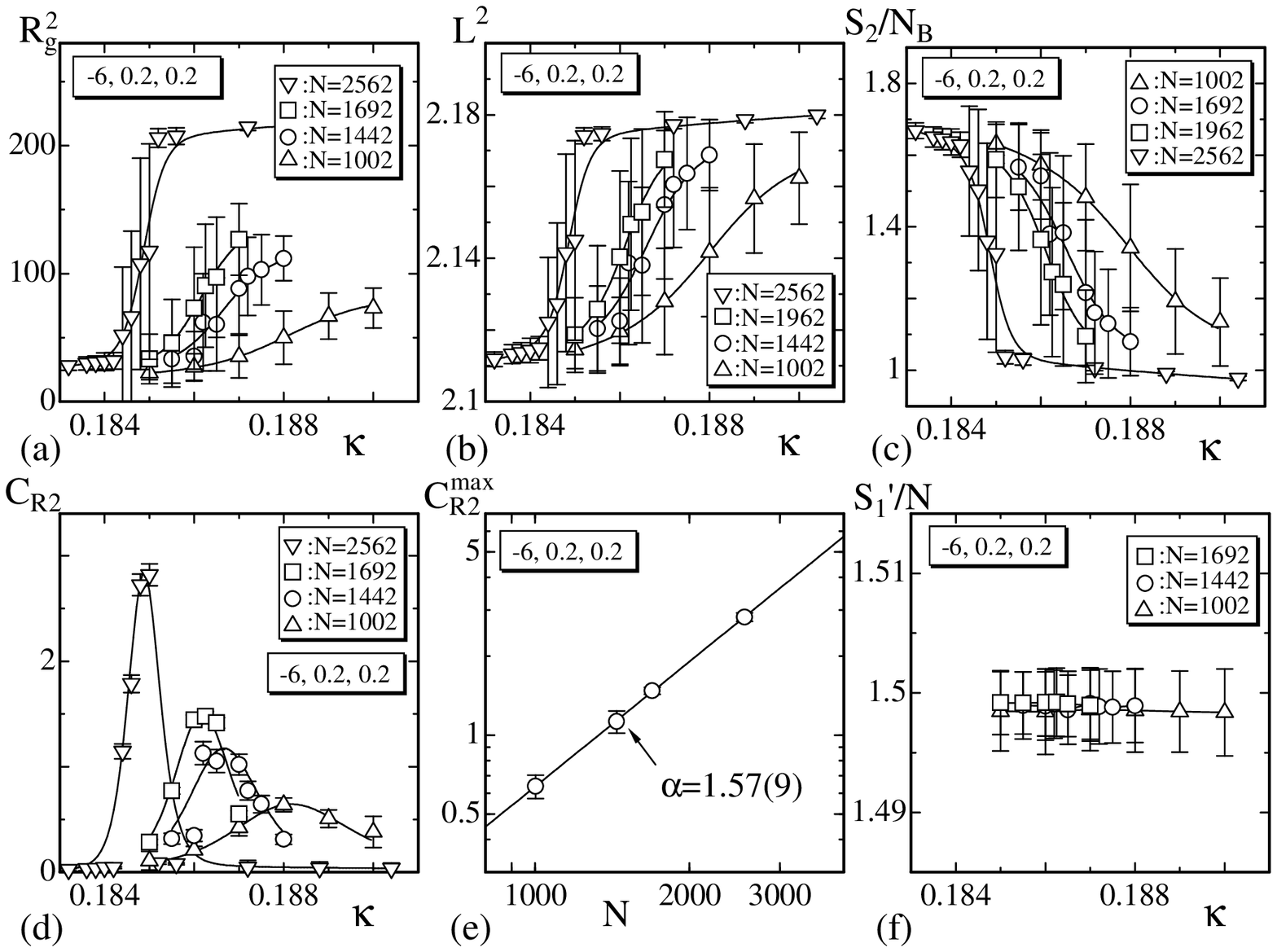}  
\caption{(a) $R_g^2$ vs. $\kappa$, (b) $L^2$ vs. $\kappa$, (c) $S_2/N_B$ vs. $\kappa$, (d) $C_{R^2}$ vs. $\kappa$, (e) the log-log plot of $C_{R^2}^{\rm max}$ vs. $N$, and (f) $S_1^\prime/N$ vs. $\kappa$. The parameters $(t,u,v)$ are fixed to  $(t,u,v)=(-6,0.2,0.2)$. The solid lines in (a)--(d) are drawn by the multi-histogram re-weighting technique.  $N_B\!=\!3N\!-\!6$ is the total number of bonds.}
\label{fig-2}
\end{figure}
Figures \ref{fig-2}(a)-\ref{fig-2}(c) show the following quantities: 1) the mean square radius of gyration $R_g^2$ defined as 
\begin{equation}
\label{X2}
R_g^2=\frac{1}{N} \sum_i \left({\bf r}_i-\bar {\bf r}\right)^2, \quad \bar {\bf r}=\frac{1}{N} \sum_i {\bf r}_i,
\end{equation}
2) the mean bond length squares $L^2$, 3) the bending energy per bond $S_2/N_B$, where the parameters are fixed to  $(t,u,v)\!=\!(-6,0.2,0.2)$ and $N_B\!=\!3N\!-\!6$ is the total number of bonds of the triangles.  We should note that $L^2\!\propto\!S_1/N_B$. From the results in Fig.2 it follows that $L^2$ discontinuously changes. It implies that the LG model has a phase transition, because the mean bond length just corresponds to the order parameter $\partial {\bf r}$ of the model as mentioned in the subsection \ref{LG-continuous}. The edge length  $\partial {\bf r}$ of triangles, or equivalently the metric ${\partial_a {\bf r}}\cdot{\partial_b {\bf r}}$, is directly connected to the shape of triangles, and therefore we expect that the transition is strongly dependent on the shape of triangles.

 Since the transition between the smooth and crumpled phases is very strong for  $(t,u,v)\!=\!(-6,0.2,0.2)$, relatively small lattices (up to $N\!=\!2562$) are sufficient to see the order of the transition. The simulations on larger lattices are meaningless if we search for a strong transition, because the data changes very sharply with varying $\kappa$ at the transition. Hence the exponents such as those in Eqs. (\ref{CX2-scaling}) and (\ref{CS2-scaling}) are hardly obtained on larger lattices. 

The variance of $R_g^2$ defined by
\begin{eqnarray}
\label{CX2}
C_{R^2}=\frac {1}{N} \left< \left(R_g^2-\langle R_g^2\rangle \right)^2 \right>,
\end{eqnarray}
and the peak values $C_{R^2}^{\rm max}$ are plotted in Figs. \ref{fig-2}(d) and \ref{fig-2}(e).
From the peak values $C_{R^2}^{\rm max}$ plotted in Fig. \ref{fig-2}(e), we have 
\begin{eqnarray}
\label{CX2-scaling}
C_{R^2}^{\rm max}\sim N^{\alpha}, \quad \alpha=1.57\pm 0.09.
\end{eqnarray}
We also obtain the variances $C_{S_2}\!=\!(\kappa^2/N) \left< \left(S_2\!-\!\langle S_2\rangle \right)^2 \right>$ and $C_{S_3}\!=\!(u^2/N) \left< \left(S_3\!-\!\langle S_3\rangle \right)^2 \right>$, and their peak values $C_{S_2}^{\rm max}$, and $C_{S_3}^{\rm max}$ are also obtained. The parameters $\kappa^2$ and $u^2$ are included in  $C_{S_2}$ and $C_{S_3}$, respectively, because both $C_{S_2}$ and $C_{S_3}$ have the meaning of a specific heat. We should note that the scaling behavior of $C_{S_2}$ and $C_{S_3}$ at the transition point is independent of whether or not the coefficients $\kappa^2$ and $u^2$ are multiplied by these quantities.  We have the following the scaling relations:
\begin{eqnarray}
\label{CS2-scaling}
C_{S_2}^{\rm max}\sim N^{\beta}, \; \beta=1.48\pm 0.03, \qquad C_{S_3}^{\rm max}\sim N^{\gamma}, \; \gamma=1.04\pm 0.05
\end{eqnarray}
The results indicate that this is the first-order transition. 
To check whether the simulation is correct or not, we plot $S_1^\prime/N$ of Eq. (\ref{S1234})
in Fig. \ref{fig-2}(f). We see that the expected result $S_1^\prime/N\!=\!3/2$ is satisfied.

\begin{figure}[!t]
\centering
\includegraphics[width=12.5cm]{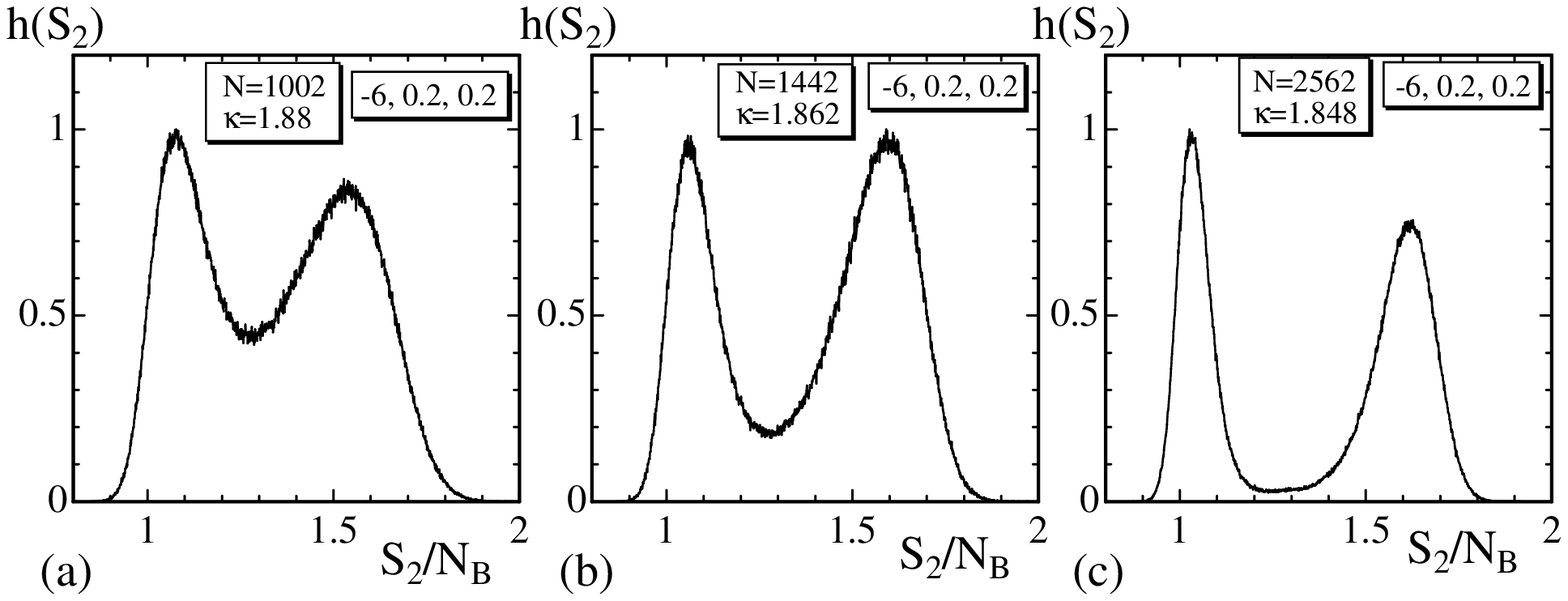}  
\caption{The normalized distribution of $S_2/N_B$ at the transition point on the lattices (a) $N\!=\!1002$, (b) $N\!=\!1442$, and (c) $N\!=\!2562$. } 
\label{fig-3}
\end{figure}
To confirm that the transition is of the first-order, we show the histogram $h(S_2)$ for the normalized distribution of  $S_2/N_B$ at the transition point (Figs. \ref{fig-3}(a)-\ref{fig-3}(c)). We see a double peak structure in $h(S_2)$ and find that the peaks become clearly separated as $N$ increases. This is typical for the first-order transition. One of the peaks corresponds to the smooth phase and the other to the crumpled phase. We should note that the double peaks can also be seen in the histograms for $S_3/N_T$ and $S_4/N_T$ just like $h(S_2)$ in Figs. \ref{fig-3}(a)-\ref{fig-3}(c).

\begin{figure}[!t]
\centering
\includegraphics[width=12.5cm]{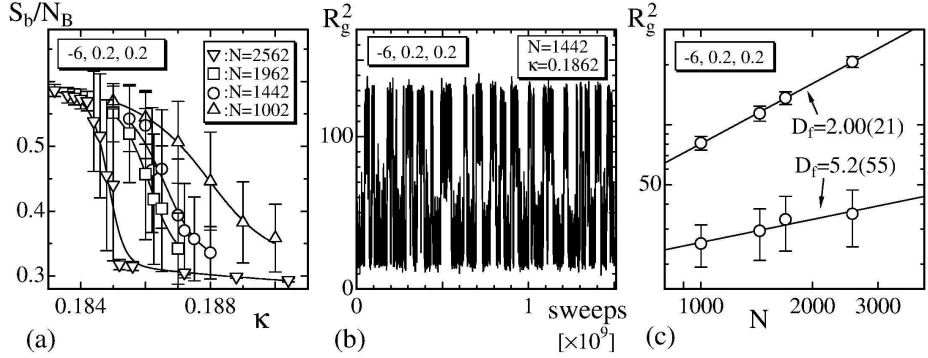}  
\caption{(a) The bending energy $S_b/N_B$ vs. $\kappa$, (b) the variation of $R_g^2$ vs. MCS, and (c) the log-log plot of $R_g^2$ vs. $N$ at the transition point, where the slope of the straight line gives the fractal dimension $D_f$. The solid lines in (a) are drawn by the multi-histogram re-weighting technique.}
\label{fig-4}
\end{figure}
We have seen that the model undergoes a first order-transition at intermediate value of $\kappa$ for $(t,u,v)\!=\!(-6,0.2,0.2)$. The smooth and crumpled phases in this transition are expected to be identical to those of the canonical model of Helfirch and Polyakov. In order to check it, the bending energy $S_b$ defined by
\begin{eqnarray}
\label{bending-energy}
S_b=\sum_{ij}\left( 1-{\bf n}_i\cdot {\bf n}_j\right) 
\end{eqnarray}
is plotted in Fig. \ref{fig-4}(a). In $S_b$, ${\bf n}_i$ is a unit normal vector of the triangle $i$, and $\sum_{ij}$ denotes the sum over all nearest neighbor triangles. We find that $S_b/N_B$ is comparable with $S_2/N_B$ of the HP model in both smooth and crumpled phases \cite{Koibuchi-PRE2005}. The variance $C_{S_b}\!=\!(1/N) \left< \left(S_b\!-\!\langle S_b\rangle \right)^2 \right>$ has a sharp peak at the transition point, and the peaks $C_{S_b}^{\rm max}$ scale according to $C_{S_b}^{\rm max}\sim N^\delta$, $\delta\!=\!1.48(3)$, which is larger than $\delta\!=\!0.93(13)$ of the canonical HP model in Ref. \refcite{Koibuchi-PRE2005}.

From the mean square radius of gyration $R_g^2$ at the transition point (Fig. \ref{fig-4}(b)), the fractal dimension $D_f$ is obtained such that
\begin{eqnarray}
\label{LG-fractal-dimension}
R_g^2 \sim N^{2/D_f}, \qquad D_f^{\rm sm}=2.00\pm0.21,\quad D_f^{\rm cr}=5.2\pm5.5,
\end{eqnarray}
where $D_f^{\rm sm}$ ($D_f^{\rm cr}$) is the fractal dimension for the smooth (crumpled) phase at the transition point  (Fig. \ref{fig-4}(c)). The result $D_f^{\rm sm}$ ($D_f^{\rm cr}$) in Eq. (\ref{LG-fractal-dimension}) is calculated from  a series of $R_g^2$  at the transition point by using only its large (small) part (Fig. \ref{fig-4}(b)). We find that$D_f^{\rm sm}\!=\!2.00(21)$ is comparable to $D_f^{\rm sm}\!=\!2.13(17)$ in Ref. \refcite{Koibuchi-PRE2004} and  $D_f^{\rm sm}\!=\!2.02(14)$ in Ref. \refcite{Koibuchi-PRE2005}. This implies that the smooth phase of the LG model is identical to those in the HP models in Refs. \refcite{Koibuchi-PRE2004,Koibuchi-PRE2005}. On the other hand, the error of $D_f^{\rm cr}\!=\!5.2(55)$ is very large, and therefore the obtained value $D_f^{\rm cr}$ is not reliable; this large error comes from the large errors of $R_g^2$ in the crumpled phase (Fig. \ref{fig-4}(c)). However, if we compare the result  $D_f^{\rm cr}$ with those of other models, we find that  $D_f^{\rm cr}\!=\!5.2(55)$ is relatively larger than $D_f^{\rm cr}\!=\!3.66(107)$ in Ref. \refcite{Koibuchi-PRE2004} and $D_f^{\rm cr}\!=\!2.59(57)$  in Ref. \refcite{Koibuchi-PRE2005}. This implies that the crumpled phase is more wrinkled than those in other models. 

The strength of the transition in the LG model varies depending on the parameters $(t,u,v)$. At $(t,u,v)\!=\!(-4,0.2,0.2)$, the transition weakens but still remains in the first order, and therefore $D_f^{\rm cr}$ is also expected to be comparable with that of the HP model if $t$ increases. As $t$ increases further, the order of the transition turns from the first order to the second or higher orders.

\subsection{Self-avoiding model and higher-dimensional model}\label{LG-SA-Highdim}
The self-avoiding (SA) interaction is able to alter the phase structure of the model \cite{KANTOR-KARDAR-NELSON-PRL1986,KANTOR-KARDAR-NELSON-PRA1987,PLISCHKE-BOAL-PRA1988,Ho-Baum-EPL1990,Grest-JPhysI1991,Kroll-Gompper-JPF1993,Munkel-Heermann-PRL1995,BCTT-PRL2001,BOWICK-TRAVESSET-EPJE2001}. The SA interaction is defined by
\begin {equation}
\label{self-avoiding}
 S_{\rm SA}=\frac{b}{2}\int d^2x d^2y \;\delta\left({\bf r}(x)-{\bf r}(y)\right),
\end{equation}
where  $b$ is the strength of self-avoidance, which corresponds to the excluded volume parameter $v$ of the Doi-Edwards model for polymers \cite{Doi-Edwards-1986}. The discrete version of $S_{\rm SA}$ is simply defined by
\begin {eqnarray}
\label{self-avoiding-dscrete}
&& S_{\rm SA}=\sum _{\triangle,\triangle'} U(\triangle,\triangle^\prime), \nonumber \\
&&  U({\it \Delta,\Delta^\prime})= \left\{
     \begin{array}{@{\,}ll}
    \infty & \; ({\rm triangles}\;{\it \Delta \Delta^\prime} \; {\rm intersect}  ), \\
             0 & \; ({\rm otherwise}).
     \end{array} 
               \right. 
\end{eqnarray}
The potential $U({\it \Delta,\Delta^\prime})$ prohibits the triangles ${\it \Delta}$ and ${\it \Delta^\prime}$ from intersecting. The parameter $b$ in the discrete $S_{\rm SA}$ is suppressed. 

\begin{figure}[!t]
\centering
\includegraphics[width=12.5cm]{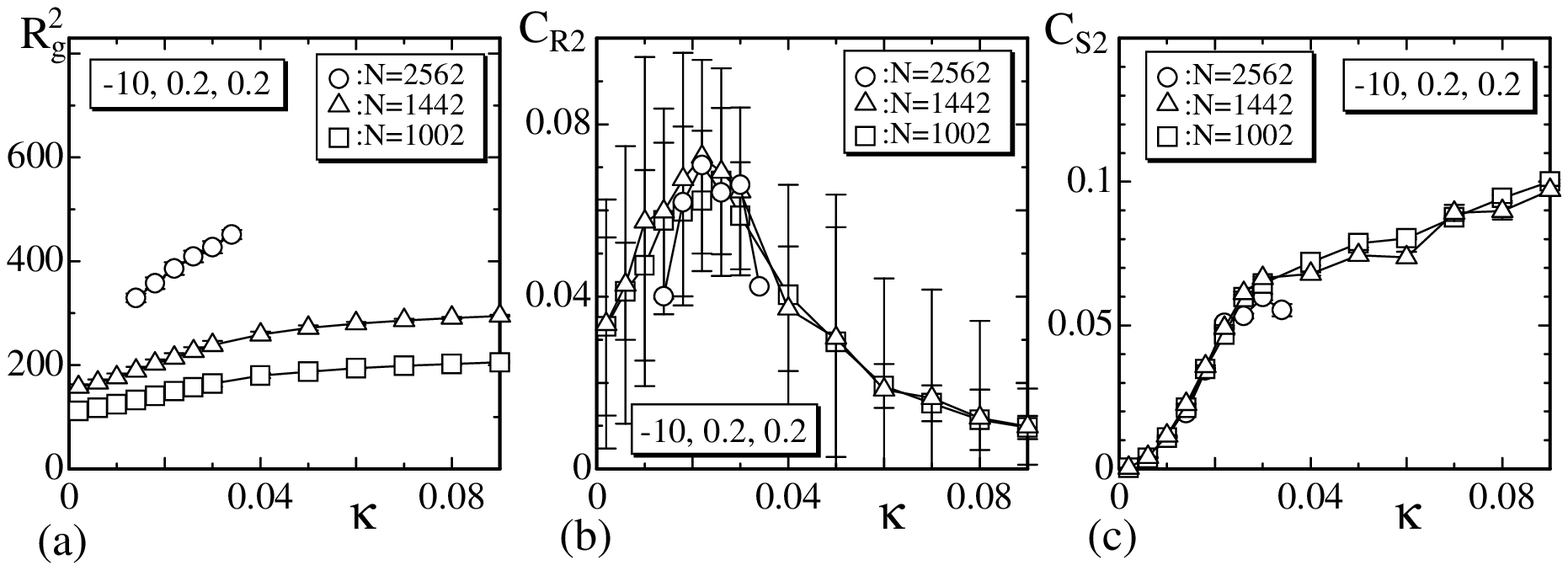} 
\caption{The MC data of the self-avoiding LG model: (a) The mean square radius of gyration $R_g^2$ vs. $\kappa$, (b) the variance $C_{R^2}$ vs. $\kappa$, and (c)  the specific heat $C_{S_2}$ vs. $\kappa$. The parameters are fixed to $(t,u,v)=(-10,0.2,0.2)$. }
\label{fig-5}
\end{figure}

The results of self-avoiding model for  $(t,u,v)\!=\!(-10,0.2,0.2)$ are shown in Figs. \ref{fig-5}(a)--(c). Recalling that the phantom model undergoes a strong first order transition under $(t,u,v)\!=\!(-10,0.2,0.2)$, we find that the transition is strongly influenced and weakened by the SA interaction. Although the variance $C_{R^2}$ has a peak, we can see no peak in the specific heat $C_{S_2}$ (Fig. \ref{fig-5}(c)). This implies that the surface fluctuation is completely suppressed by the SA interaction in the case of the LG model.

\begin{figure}[!t]
\centering
\includegraphics[width=12.5cm]{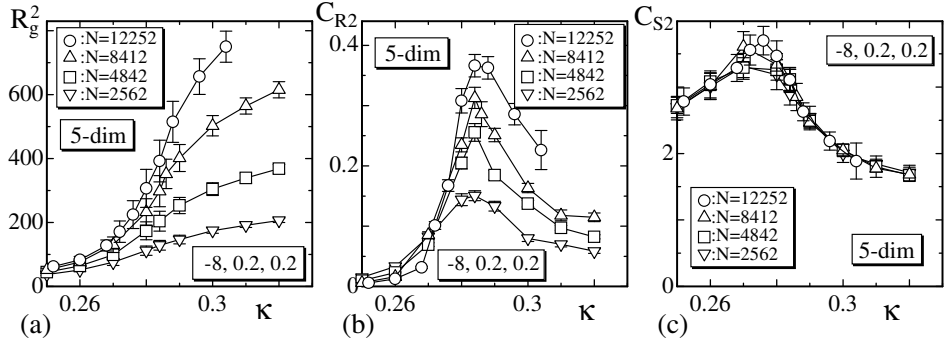}  
\caption{The MC data of the LG model in ${\bf R}^5$: (a) The mean square radius of gyration $R_g^2$ vs. $\kappa$, (b) the variance $C_{R^2}$ vs. $\kappa$, and (c) the specific heat $C_{S_2}$ vs. $\kappa$. The parameters are fixed to $(t,u,v)\!=\!(-8,0.2,0.2)$.}
\label{fig-6}
\end{figure}
The transition is also expected to be weakened in the higher-dimensional spaces ${\bf R}^d  (d\!>\!3)$. Indeed, we see that $R_g^2$ smoothly varies against $\kappa$ in ${\bf R}^5$ (Fig. \ref{fig-6}(a)) and the transition weakens even for $(t,u,v)\!=\!(-8,0.2,0.2)$, for which the model undergoes a strong first-order transition in ${\bf R}^3$. From the variance plot of $C_{R^2}$ vs. $\kappa$ in Fig. \ref{fig-6}(b) it is clearly seen that the surface size rapidly changes during the transition just like the first-order transition case. However, the specific heat $C_{S_2}$ remains unchanged if $N$ increases (Fig. \ref{fig-6}(c)), and therefore, the order of the transition is considered to be weakened to a higher-order one. The simulations are performed in ${\bf R}^{10}$ for $(t,u,v)\!=\!(-8,0.2,0.2)$, and we see that the transition is further weakened as compared to the one in ${\bf R}^5$.

\section{David-Guitter model for membranes}\label{DG-model}
\subsection{Continuous Hamiltonian}\label{DG-continuous}
Our next goal is to investigate  numerically whether or not the surface model of F.David and E.Guitter (DG)  undergoes a discontinuous transition between the smooth phase at $\kappa\!\to\! \infty$ and the collapsed phase at  $\kappa\!\to\! 0$, where $\kappa$ is the bending rigidity. The bending energy of the DG model is identical to the one of the LG model.

The Hamiltonian $S_{\rm DG}$ of the model is given by \cite{David-Guitter-1988EPL,FDavid-SMMS2004}
\begin {eqnarray}
\label{DG-Hamiltonian}
&&S_{\rm DG}=\kappa S_2+ S_5,  \nonumber \\
&&S_2=\frac{1}{2}\int d^2x \left({\partial ^2 {\bf r}}\right)^2,\\
&& S_5=\mu S_{51}+\lambda S_{52}=\int d^2x \left(\mu u_{ab} u_{ab} + \frac{\lambda}{2} u_{aa} u_{bb}\right), \nonumber
\end{eqnarray}
where $u_{ab}$ is the strain tensor defined by
\begin{equation}
\label{strain}
u_{ab}=\frac{1}{2}\left(\frac{\partial {\bf r}}{\partial x_a}\cdot \frac{\partial {\bf r}}{\partial x_b}-\delta_{ab}\right).
\end{equation}
We use the symbol $S_5$ for the second term of the Hamiltonian $S_{\rm DG}$ to distinguish it from $S_3$ and $S_4$ of the LG model in the previous section.

The surface size increases infinitely if $\mu\!=\!\lambda\!=\!0$, because $S_2$ imposes no constraint on the surface size while the entropy increases with increasing size. Therefore, we assume $\mu\!\not=\!0$ and $\!\lambda\!\not=\!0$. Moreover, we concentrate on the small $\kappa$ region, because the surface is expected be always smooth at sufficiently large $\kappa$ where no transition is expected. 

The strain tensor $u_{ab}$ in Eq. (\ref{strain}) is considered to be a deviation of the induced metric ${\partial_a {\bf r}}\cdot{\partial_b {\bf r}}$ from the Euclidean metric $\delta_{ab}$. The reason for introducing such variable $u_{ab}$ is because ${\partial_a {\bf r}}\cdot{\partial_b {\bf r}}$ becomes proportional to $\delta_{ab}$ when the LG Hamiltonian has a double minima for $t\!<\!0$ in Eq. (\ref{disc_LG}) in the mean field approximation. In the case for $t\!<\!0$, small (large) $u_{ab}$ corresponds to the condition that the triangles are almost regular with constant edge length (not always regular and uniform) on the triangulated surfaces. 

In addition, the transition is expected to be influenced, actually strengthened, on the surfaces composed of regular triangles as mentioned in the Introduction. This implies that the transition is in close relation to the regularity of triangles. It is possible that the transition is seen only if the triangles are uniform in size, and therefore this should be checked numerically. This is the main reason why we study the DG model. In fact, we have  ${\partial_a {\bf r}}\cdot{\partial_b {\bf r}}\!=\!\delta_{ab}$ only on the regular square lattice of uniform size, and  ${\partial_a {\bf r}}\cdot{\partial_b {\bf r}}$ can only be close to $\delta_{ab}$ on the regular triangle lattice of constant edge length.  This implies that the energies $S_{51}$ and $S_{52}$ become minimal in the limit of $u_{ab}\to 0$. We should note that ${\partial_a {\bf r}}\cdot{\partial_b {\bf r}}$ is not exactly equal to $\delta_{ab}$ even on the uniform and regular triangles because ${\partial_a {\bf r}}\cdot{\partial_b {\bf r}}\!\not=\!0 (a\!\not=\!b)$ there. Nevertheless, we can understand that the transition is related to the homogeneity of the lattice because the transition occurs only on almost regular triangle lattices.

The terms of $u_{ab}u_{ab}$ and  $u_{aa}u_{bb}$  in $S_5$ are given by 
\begin{eqnarray}
\mu S_{51}&=&\mu\int d^2xu_{ab}u_{ab}= \mu\int d^2x\left(u_{11}^2+u_{22}^2+u_{12}^2+u_{21}^2\right)  \nonumber\\
 &=&\frac{\mu}{4}\int d^2x\left(\left[\left({\partial_1 {\bf r}}\right)^2\!-\!1\right]^2+\left[\left({\partial_2 {\bf r}}\right)^2\!-\!1\right]^2  
+2\left[{\partial_1 {\bf r}}\cdot{\partial_2 {\bf r}}\right]^2\right) 
\end{eqnarray}
and 
\begin{eqnarray}
&&\lambda S_{52}=\frac{\lambda}{2}\int d^2x u_{aa}u_{bb}=\frac{\lambda}{2}\int d^2x  \left(u_{11}+u_{22}\right)^2 \nonumber\\
&=&\frac{\lambda}{2}\int d^2x \left(u_{11}^2+u_{22}^2+2u_{11}u_{22}\right) \\
&=&\frac{\lambda}{8}\int d^2x \left(\left[\left({\partial_1 {\bf r}}\right)^2-1\right]^2+\left[\left({\partial_2 {\bf r}}\right)^2-1\right]^2 +2\left[\left({\partial_1 {\bf r}}\right)^2-1\right]\left[\left({\partial_2 {\bf r}}\right)^2-1 \right]\right). \nonumber
\end{eqnarray}

The parameters $\mu$ and $\lambda$ are called the Lame coefficients.

\subsection{Discrete Hamiltonian}\label{DG-discrete}
The bending energy $S_2$ is identical to that of the LG model in Eq. (\ref{LG-Hamiltonian}). Thus, the discrete expressions of $S_2$ and $S_5$ are given by
\begin{eqnarray}
\label{disc_DG_1}
S_{\rm DG}&=&\kappa S_2 + S_5,  \nonumber \\
S_2&=&\frac{1}{3}\sum_{ij}\left({\bf e}_i- {\bf e}_j\right)^2+\frac{1}{3}\sum_{(ij),(kl)}\left({\bf e}_i- {\bf e}_j\right)\cdot\left({\bf e}_k- {\bf e}_l\right), \nonumber \\
S_5&=&\mu S_{51}+\lambda S_{52}  \nonumber \\
&=&\frac{\mu}{6} \sum_{i=1}^{N_T} \left[\sum_{j=1}^{3}\left({\bf e}_{j}^2-1\right)^2+\left({\bf e}_1\cdot{\bf e}_2\right)^2+\left({\bf e}_2\cdot{\bf e}_3\right)^2+\left({\bf e}_3\cdot{\bf e}_1\right)^2\right]  \nonumber \\ 
&&+\frac{\lambda}{12}\sum_{i=1}^{N_T}\left[\sum_{j=1}^{3}\left({\bf e}_{j}^2-1\right)^2+ \left({\bf e}_1^2-1\right)\left({\bf e}_2^2-1\right)  \right.  \nonumber\\
&&\hspace{1.8cm} 
+\left({\bf e}_2^2-1\right)\left({\bf e}_3^2-1\right)+ \left({\bf e}_3^2-1\right)\left({\bf e}_1^2-1\right)\Biggr]. 
\end{eqnarray} 

$\sum_{i=1}^{N_T}$ in $S_5(=\!\mu S_{51}\!+\!\lambda S_{52})$  is the sum over all triangles $i$, where $N_T$ is the total number of triangles, and $\sum_{j}$ is the sum of edge vectors of the triangle $i$. 
It is easy to see that $S_5$ in Eq. (\ref{disc_DG_1}) is a discretization of continuous $S_5$ in Eq. (\ref{DG-Hamiltonian}). Note that the continuous $S_5$  in Eq. (\ref{DG-Hamiltonian}) includes only the first order derivatives of ${\bf r}$, and for this reason $S_5$ can be discretized on a triangle. Unlike $S_5$, the sum $S_2$ in Eq. (\ref{disc_DG_1}) contains the second order derivatives of ${\bf r}$ (= Laplacian)  and thus several neighboring triangles are necessary for proper discretization of $S_2$.

The Lame coefficients $\mu$ and $\lambda$ control both the size and the shape of triangles; the regular triangle of constant edge length is expected in the limit of $\mu\!\to\! \infty$ and  $\lambda\!\to\! \infty$ on triangulated surfaces.

\subsection{Monte Carlo results}\label{DG-Monte}

\begin{figure}[!t]
\centering
\includegraphics[width=12.5cm]{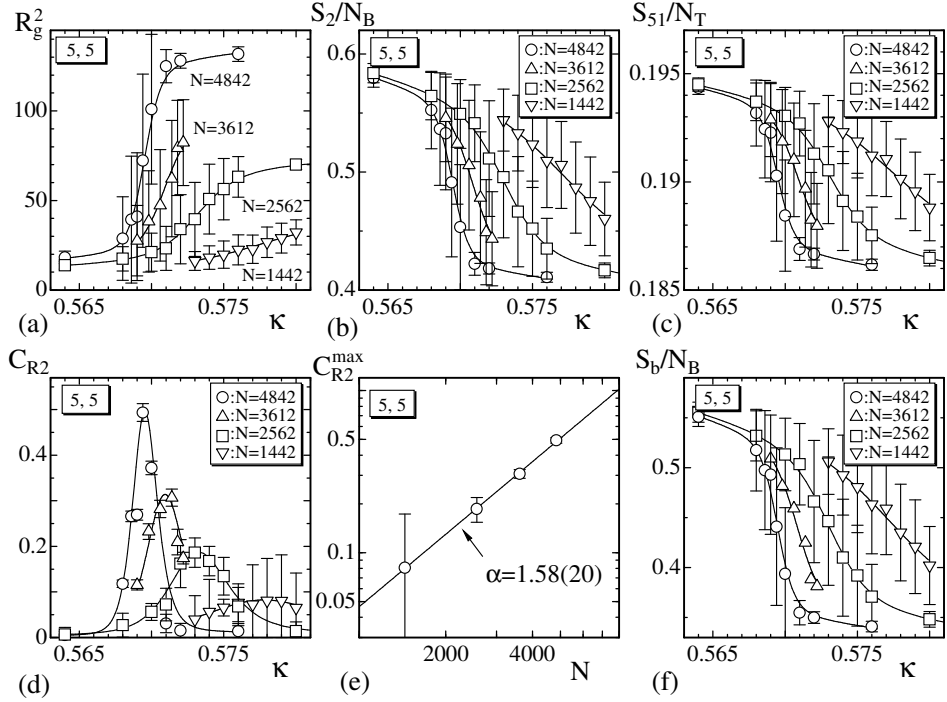}  
\caption{The MC data of the DG model: (a) $R_g^2$ vs. $\kappa$, (b) $S_2/N_B$ vs. $\kappa$, (c) $S_{51}/N_T$ vs. $\kappa$, (d) $C_{R^2}$ vs. $\kappa$, (e) the log-log plot of $C_{R^2}^{\rm max}$ vs. $N$, and (f) $S_b/N_B$ vs. $\kappa$. The parameters $(\mu,\lambda)$ are fixed to  $(\mu,\lambda)\!=\!(5,5)$. The solid lines except the one in (e) are drawn by the multi-histogram re-weighting technique.  }
\label{fig-7}
\end{figure}
 We firstly fix the parameters $\mu$ and $\lambda$ such that $(\mu,\lambda)\!=\!(5,5)$. The mean square radius of gyration $R_g^2$, the bending energy per bond $S_2/N_B$, the energy $S_{51}$ per triangle $S_{51}/N_T$
  plotted in Figs. \ref{fig-7}(a)--\ref{fig-7}(c) have large errors just like those in Figs.\ref{fig-2}(a)--\ref{fig-2}(c). This indicates that the transition is of the first order. The fact that the peak value of the variances $C_{R^2}^{\rm max}$ grows with increasing of $N$ confirms it - see Fig.\ref{fig-7}(d). The log-log plot of $C_{R^2}^{\rm max}$ vs. $N$ in Fig.\ref{fig-7}(e) gives a scaling relation such that $C_{R^2}^{\rm max}\!\sim\! N^\alpha, \alpha\!=\!1.58(20)$. We have also the relations $C_{S_2}^{\rm max}\!\sim\! N^{\beta}, \beta\!=\!1.52(12)$ and $C_{S_{51}}^{\rm max}\!\sim\! N^{\gamma}, \gamma=1.32(26)$. These data imply that the transition is of the first order. The variation of the bending energy $S_b/N_B$ of Eq. (\ref{bending-energy}) against $\kappa$ in Fig. \ref{fig-7}(f) is comparable to that of $S_2/N_B$ and $S_{51}/N_T$. The value of $S_b/N_B$ is almost identical to  that of $S_b/N_B$ in Fig. \ref{fig-4}(a) of the LG model. It implies that the surface smoothness at the transition is almost identical to that of the LG model at the transition. 

\begin{figure}[!t]
\centering
\includegraphics[width=12.5cm]{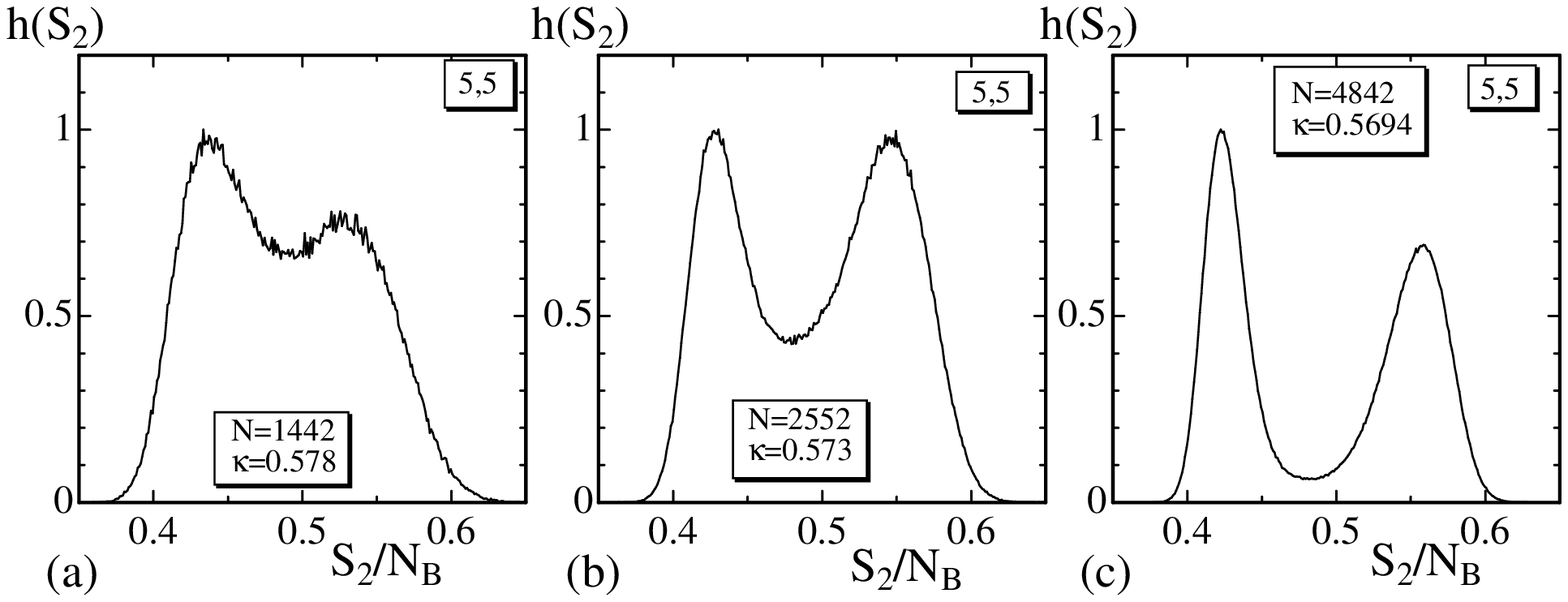}  
\caption{The normalized distribution of $S_2/N_B$ at the transition point on the lattices (a) $N\!=\!1442$, (b) $N\!=\!2562$, and (c) $N\!=\!4842$. } 
\label{fig-8}
\end{figure}
To confirm that the transition is of the first order, we show the normalized distribution $h(S_2)$ of $S_2/N_B$ at the transition point in Fig. \ref{fig-8}. The double peak structure in $h(S_2)$ is a signal of the first order transition. Indeed, the peaks become clear as $N$ increases.    

\begin{figure}[!t]
\centering
\includegraphics[width=12.5cm]{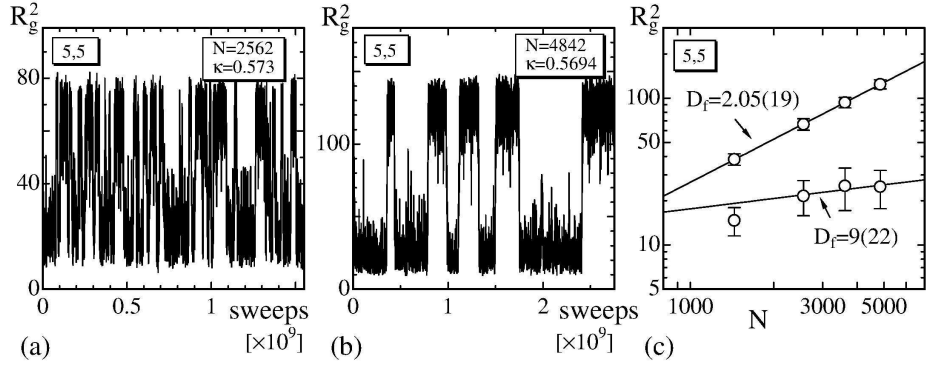}  
\caption{(a) The bending energy $S_b/N_B$ vs. $\kappa$, (b) the variation of $R_g^2$ vs. MCS, and (c) the log-log plot of $R_g^2$ vs. $N$ at the transition point, where the slope of the straight line gives the fractal dimension $D_f$. }
\label{fig-9}
\end{figure}
Coexistence of two different states at the transition point is apparent in the variation of $R_g^2$ vs. MCS (Figs.  \ref{fig-9}(a),(b)). Computing the average of $R_g^2$ in the smooth and crumpled phases separately, we have the fractal dimension $D_f$ via $R_g^2 \sim N^{2/D_f}$. The results are
\begin{eqnarray}
\label{DG-fractal-dimension}
D_f^{\rm sm}=2.05\pm0.19,\quad D_f^{\rm cr}=9.0\pm2.2,
\end{eqnarray}
in which $D_f^{\rm sm}$ ($D_f^{\rm cr}$) is almost the same as (relatively larger than) that of the LG model in Eq. (\ref{LG-fractal-dimension}). This also indicates that the transition is a quite strong first order one in the DG model for $(\mu,\lambda)\!=\!(5,5)$.

We discuss the phase transition dependence on the surface metric degrees of freedom. Numerical data confirm that the transition weakens when the parameters $(\mu,\lambda)$ become smaller such that $(\mu,\lambda)\!=\!(3,3)$, and eventually the transition disappears with decreasing of $(\mu,\lambda)$. This implies that the metric degrees of freedom play an important role in the transition. Indeed, since ${\partial_a {\bf r}}\cdot{\partial_b {\bf r}}$ is the induced metric, the strain tensor $u_{ab}$ measures the distance between the induced metric ${\partial_a {\bf r}}\cdot{\partial_b {\bf r}}$ and the Euclidean metric $\delta_{ab}$. The numerical results show that the phase transition can be seen only when this distance is small. Intuitively, the condition that ${\partial_a {\bf r}}\cdot{\partial_b {\bf r}}$ is close to $\delta_{ab}$ makes the triangles almost regular and uniform in size in the discrete model. 

\begin{figure}[!t]
\centering
\includegraphics[width=12.5cm]{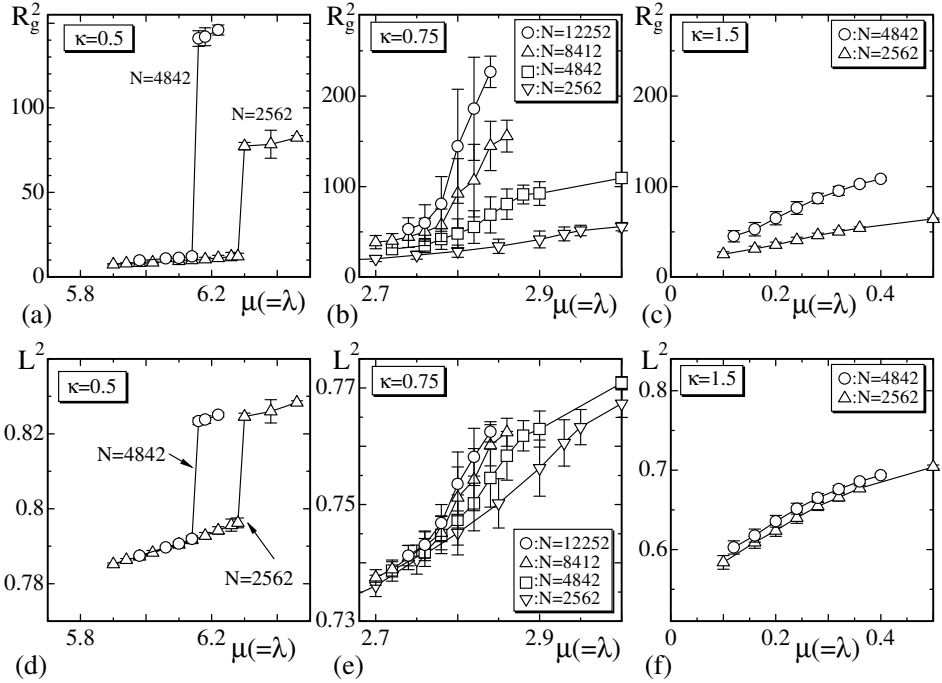}  
\caption{The mean square radius of gyration $R_g^2$ vs. $\mu(=\lambda)$  for (a) $\kappa\!=\!0.5$,  (b) $\kappa\!=\!0.75$  and (c) $\kappa\!=\!1.5$, and the mean bond length squares $L^2$ vs. $\mu(=\lambda)$  for (d) $\kappa\!=\!0.5$,  (e) $\kappa\!=\!0.75$  and (f) $\kappa\!=\!1.5$.}
\label{fig-10}
\end{figure}
  We have seen that the model undergoes the transition at the low bending region only when $(\mu,\lambda)$ are sufficiently large, where the surfaces are uniform in the sense that the triangles are regular and uniform in size. 
 However, the reason why the transition disappears with decreasing $(\mu,\lambda)$ still remains to be clarified.  
To see this, we plot the MC data in Fig. \ref{fig-10}, where $(\mu,\lambda)$ are varied while $\kappa$ is fixed to $\kappa\!=\!0.5$,  $\kappa\!=\!0.75$ and  $\kappa\!=\!1.5$.  We see from $R_g^2$ in Fig. \ref{fig-10}(a) that the model has a strong first-order transition at $(\mu,\lambda)\!\simeq\! 6$ for $\kappa\!=\!0.5$. The smooth phase  at $(\mu,\lambda)\!>\! (6,6)$ is separated  from the collapsed phase at $(\mu,\lambda)\!<\! (6,6)$ by the first order transition. The transition still remains first order for $\kappa\!=\!0.75$ (Fig. \ref{fig-10}(b)), and it eventually weakens with increasing $\kappa$ and disappears when $\kappa$ increases to $\kappa\!=\!1.5$ (Fig. \ref{fig-10}(c)). The mean bond length squares $L^2$ vs. $\mu(=\lambda)$ also changes almost identical with $R_g^2$ as $\kappa$ increases (Figs. \ref{fig-10}(d)--(f)). Moreover the peaks of the specific heats $C_{S_2}$, $C_{S_{51}}$ and $C_{S_{52}}$, which are not shown, remain constant even when the surface size increases for $\kappa\!=\!0.5$. These results show that the smooth phase emerges only when the triangles are sufficiently uniform both in size and shape. Thus, we find that the transition disappears with decreasing $(\mu,\lambda)$ such that $(\mu,\lambda)\!\to\! (0,0)$; there is no smooth phase at $(\mu,\lambda)\!\to\! (0,0)$ for the low and intermediate bending region. 

The result that the transition can only be seen on the uniform triangle lattices is consistent with the results of the meshwork model in which the transition is strengthened on the uniform lattices \cite{Endo-Koibuchi-EPJB2008}. In the case of the meshwork model, the edge length of triangles is constrained to be constant due to the scale invariance of the partition function $Z$, and for this reason the triangles always become almost homogeneous in contrast to the DG model. Moreover, the meshwork becomes further homogeneous due to the presence of the in-plane shear energy $S_3\!=\!\sum_{i}\left[1\!-\!\cos(\theta_{i}\!-\!\pi/3)\right]$ in the Hamiltonian, and as a consequence the transition is strengthened. This is also true for the DG (and LG) model, although the scale invariance of $Z$ does not make the bond length constant.

\section{Summary and conclusion}\label{Conclusion}
We have numerically studied the Landau-Ginzburg (LG) and David-Guitter (DG) models for membranes. Both models undergo the first order transition between the collapsed and smooth phases. We find that the LG and DG models do not contradict the Helfrich-Polyakov (HP) model as far as the order of the transition is concerned. It is also found that the transition of the LG model weakens on the self-avoiding surfaces and in the higher dimensional spaces in accordance with the expectation. 

The dependence of the transition on the surface metric is studied in the DG model. The distance between the Euclidean metric $\delta_{ab}$ and the induced metric ${\partial_a {\bf r}}\cdot{\partial_b {\bf r}}$ is implemented via the strain tensor ${\partial_a {\bf r}}\cdot{\partial_b {\bf r}}\!-\!\delta_{ab}$ in the DG model. Numerical data indicate that the transition can be seen only when ${\partial_a {\bf r}}\cdot{\partial_b {\bf r}}$ is sufficiently close to $\delta_{ab}$, which happens on the homogeneous surfaces because they are composed of almost regular triangles with constant edge length. The obtained results imply that only homogeneous surfaces undergo the transition between the collapsed and smooth phases.





\begin{thebibliography}{00}
\bibitem{PKN-PRL1988}
M. Paczuski, M. Kardar and D.R. Nelson, Phys. Rev. Lett. {\bf 60},  2638 (1988).

\bibitem{Bowick-SMMS2004}
M.J. Bowick, {\it Fixed-connectivity Membranes} in {\it Statistical Mechanics of Membranes and Surfaces, Second Edition}, edited by  D. Nelson, T. Piran, and S. Weinberg, (World Scientific, 2004) p.323. 

\bibitem{HELFRICH-1973}
 W. Helfrich, Z. Naturforsch {\bf 28c},  693 (1973).

\bibitem{POLYAKOV-NPB1986}
 A.M. Polyakov, Nucl. Phys. B {\bf 268},  406 (1986).

\bibitem{KANTOR-NELSON-PRA1987}
Y. Kantor and D.R. Nelson, Phys. Rev. A {\bf 36}, 4020 (1987). 

\bibitem{Kantor-SMMS2004}
Y. Kantor, {\it Properties of Thethered Surfaces} in {\it Statistical Mechanics of Membranes and Surfaces, Second Edition}, edited by  D. Nelson, T. Piran, and S. Weinberg, (World Scientific, 2004) p.111. 

\bibitem{Gompper-Kroll-SMMS2004}
G. Gompper and D.M. Kroll, {\it Triangulated-surface Models of Fluctuating Membranes} in {\it Statistical Mechanics of Membranes and Surfaces, Second Edition}, edited by  D. Nelson, T. Piran, and S. Weinberg, (World Scientific, 2004) p.359. 

\bibitem{Endo-Koibuchi-EPJB2008}
I. Endo and H. Koibuchi, Euro. Phys. J. B {\bf 66}, 467 (2008). 

\bibitem{Koibuchi-EPJB2007}
H. Koibuchi, Euro. Phys. J. B {\bf 59}, 55 (2007). 

\bibitem{David-Guitter-1988EPL}
F. David and E. Guitter, Europhys. Lett.  {\bf 5}, 709 (1988).

\bibitem{FDavid-SMMS2004}
F. David, {\it Geometry and Field Theory of Random Surfaces and Membranes} in {\it Statistical Mechanics of Membranes and Surfaces, Second Edition}, edited by  D. Nelson, T. Piran, and S. Weinberg, (World Scientific, 2004) p. 149. 

\bibitem{KD-PRE2002}
J. -P. Kownacki  and H.T. Diep,  Phys. Rev. E {\bf 66}, 066105 (2002).

\bibitem{Koibuchi-PRE2004}
H. Koibuchi, N. Kusano, A. Nidaira, K. Suzuki, and M. Yamada Phy. Rev. E {\bf 69}, 066139(1-6) (2004). 

\bibitem{Koibuchi-PRE2005}
H. Koibuchi and T. Kuwahata, Phys. Rev. E {\bf 72}, 026124(1-6) (2005).

\bibitem{WHEATER-JP1994}
J.F. Wheater, J. Phys. A Math. Gen. {\bf 27}, 3323 (1994). 


\bibitem{KANTOR-KARDAR-NELSON-PRL1986}
Y. Kantor, M. Kardar and D.R. Nelson,  Phys. Rev. Lett. {\bf 57}, 791 (1986). 

\bibitem{KANTOR-KARDAR-NELSON-PRA1987}
Y. Kantor, M. Kardar and D.R. Nelson,  Phys. Rev. A {\bf 35}, 3056 (1987). 

\bibitem{PLISCHKE-BOAL-PRA1988}
M. Plischke and D. Boal,  Phys. Rev. A {\bf 38}, 4943 (1988). 

\bibitem{Ho-Baum-EPL1990}
J. -S. Ho  and A. Baumg${\ddot {\rm a}}$rtner,  Europhys. Lett. {\bf 12}, 295 (1990).

\bibitem{Grest-JPhysI1991}
G. Grest, J. Phys. I (France),  {\bf 1}, 1695 (1991).

\bibitem{Kroll-Gompper-JPF1993}
D.M. Kroll and G. Gompper,  J. Phys. I. France  {\bf 3}, 1131 (1993). 

\bibitem{Munkel-Heermann-PRL1995}
C. M${\ddot {\rm u}}$nkel and D.W. Heermann,  Phys. Rev. Lett. {\bf 75}, 1666 (1995). 

\bibitem{BCTT-PRL2001}
M. Bowick, A. Cacciuto, G. Thorleifsson, and A. Travesset,  Phys. Rev. Lett. \textbf{87}, 148103 (2001).

\bibitem{BOWICK-TRAVESSET-EPJE2001}
M. Bowick, A. Cacciuto, G. Thorleifsson, and A. Travesset,  Euro. Phys. J. E \textbf{5}, 149 (2001). 

\bibitem{Doi-Edwards-1986}
M. Doi and F. Edwards, {\it The Theory of Polymer Dynamics}, (Oxford University Press, 1986). 



\end{thebibliography}
\end{document}